\documentclass{elsart1p}
\usepackage{graphicx}
\usepackage{amssymb}
\def \GX    {{\textsc{GlueX}}}
\begin{document}
\begin{frontmatter}
%
%
%
\title{The 12 GeV JLab Upgrade Project}
%
%
\author{Elton S. Smith\corauthref{cor1}}
\corauth[cor1]{E-mail address: elton@jlab.org.}

\address{Thomas Jefferson National Accelerator Facility, Newport News, Virginia 23606 USA}

\begin{abstract}
The upgrade of the CEBAF Accelerator at Jefferson Lab to 12 GeV will
deliver high luminosity and high quality beams, which will open unique
opportunities for studies of the quark and gluon structure of hadrons in the valence region. Such
physics will be made accessible by substantial additions to the
experimental equipment in combination with the increased energy reach of
the upgraded machine. The emphasis of the talk will be on the program in a
new experimental Hall D designed to search for gluonic excitations.
\end{abstract}
\end{frontmatter}
%

\section{Overview of the project}
The 12 GeV Upgrade presents a unique opportunity  for the nuclear physics community to expand its reaches 
into unknown scientific areas and allow researchers to probe the quark and gluon structure 
of strongly interacting systems.
We first review the scope of the project. Then, due to limited space, we only feature two physics programs
(study of gluonic excitations and study of Generalized Parton Distributions) and conclude with 
the current status of project construction. The description of the full experimental program
using the upgraded machine can be found in 
Ref.\,\cite{12gevCDR}.

\begin{figure}[t]
\begin{center}
\includegraphics[width=7cm]{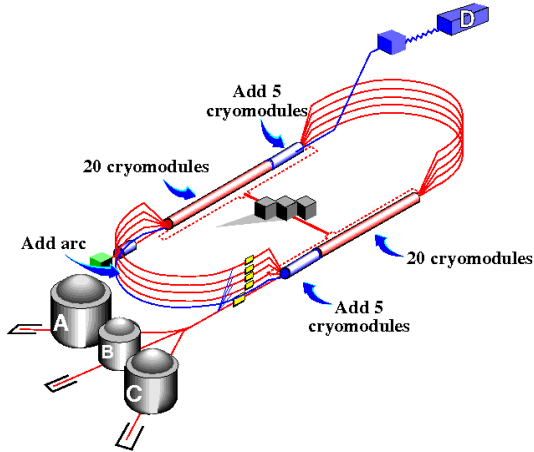}
\caption{\label{accelerator} Layout of the Jefferson Lab CEBAF accelerator indicating additions needed
for the 12 GeV Upgrade Project.}
\end{center}
\end{figure}

The scope of the JLab 12 GeV Upgrade Project includes doubling the present energy of
the accelerator, major enhancements to the equipment in
the existing experimental areas, and the
construction of a new experimental area (Hall D) with a new detector. 
In order to support these additional facilities, the project also covers
the civil construction of the buildings and infrastructure associated with the Hall D complex and a major
addition to the Central Helium Liquefier (CHL) to double the cryogenic capacity for the
accelerator upgrade.  

The present CEBAF accelerator (Fig.\,\ref{accelerator}) recirculates electrons
through two superconducting linacs between one and five times delivering simultaneous independent 
beams to the current three experimental areas (A, B and C). The maximum gain per pass is 1.2 GeV.
The typical beam energies on the experimental targets range from 0.8 to 5.7 GeV with
100\% duty cycle, maximum total current of 200 $\mu$A, and 75\% polarization. Fortunately,
the present footprint of the machine allows for growth. The 1.4 km 
racetrack tunnel is large enough for the magnetic arcs to be able to 
accommodate a 12-GeV electron beam, and 20\% of the linac sections are
empty allowing additional accelerating cavities to be installed. The new 
cryomodule design aims to exceed the original CEBAF specification by a factor of five.  
The upgrade configuration will result in a maximum energy
gain per pass of 2.2 GeV, providing the existing Halls A, B and C with up to 11 GeV.
The maximum energy to Hall D will be 12 GeV, which will be achieved by adding
a tenth arc and recirculating the beam 5.5 times before delivery. 
Three independent polarized beams can be delivered to the experimental areas
with a current of up to 90$\mu$A at the maximum beam energy.

\begin{figure}[t]
\begin{minipage}[t]{5.5cm}
\begin{center}
\includegraphics[width=5cm]{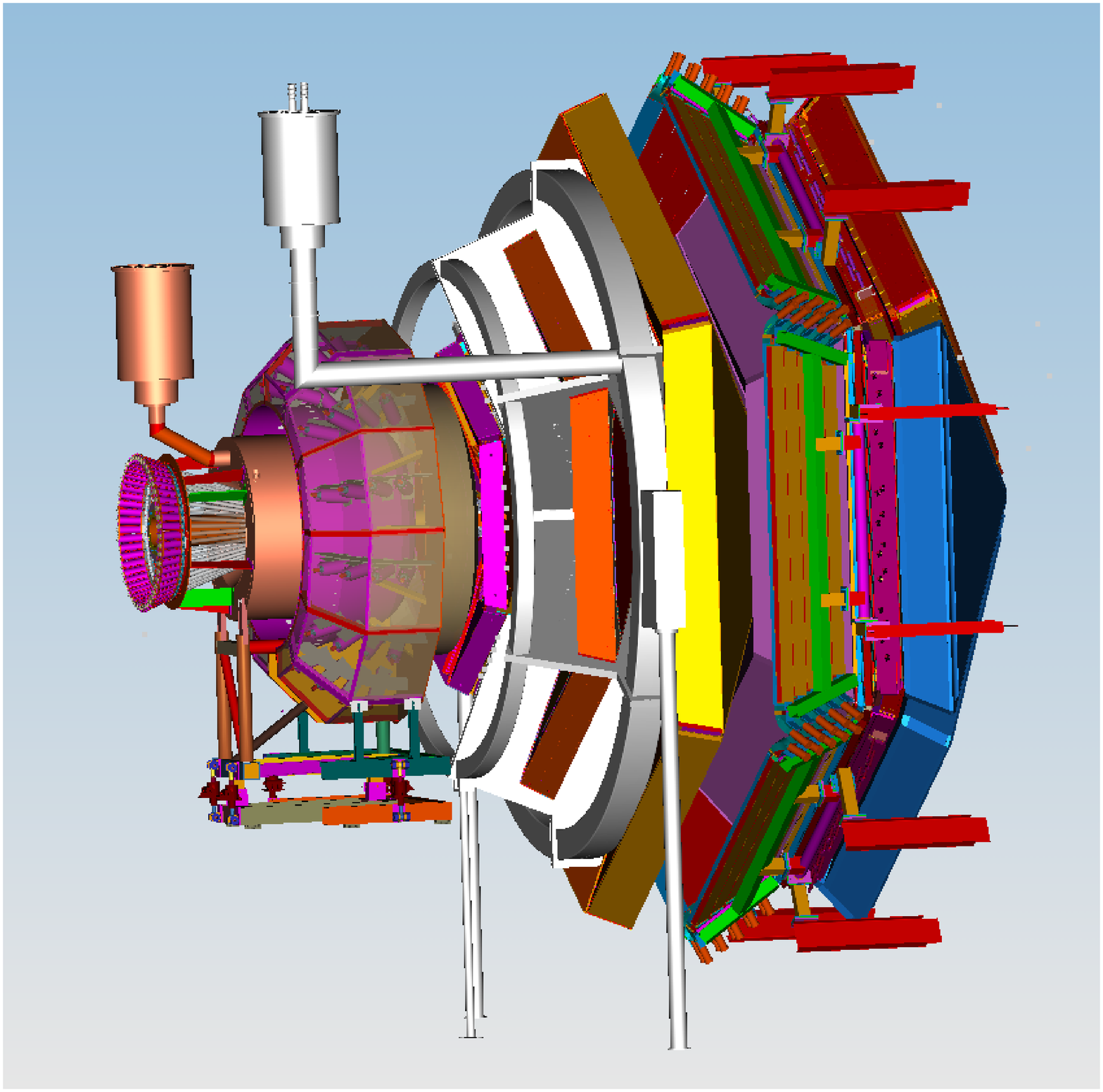}
\caption{\label{CD3_CLAS12} CLAS12 detector in Hall B.}
\end{center}
\end{minipage}
\hfill
\begin{minipage}[t]{7.5cm}
\begin{center}
\includegraphics[width=7cm,bb=0 0 830 570]{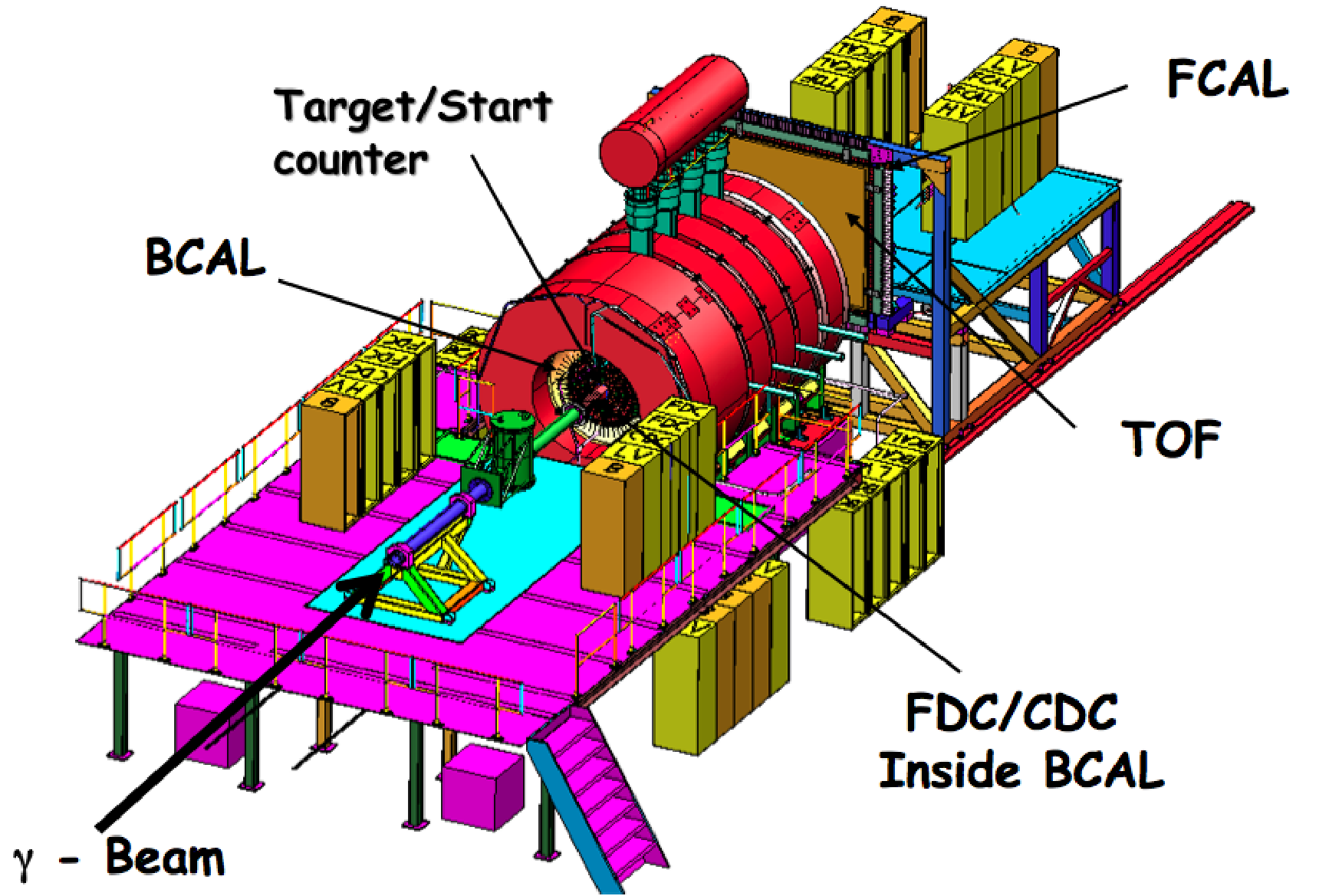}
\caption{\label{hddetector_platform} Layout of the Hall D detector and electronics.}
\end{center}
\end{minipage}
\end{figure}
\begin{figure}[t]
\begin{center}
\includegraphics[width=14cm,bb=0 0 1330 485]{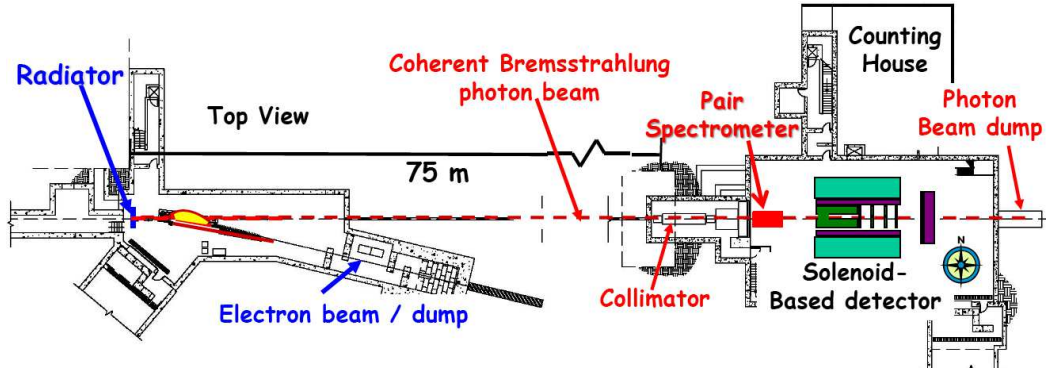}
\caption{\label{Photon_beam_layout} Layout of the Hall D photon beamline.}
\end{center}
\end{figure}

In order to capitalize on the physics opportunities offered by the increased kinematic reach
and quality of beams of the upgraded accelerator, the experimental equipment in each of the
halls will have significant improvements. 
The CEBAF Large Acceptance Spectrometer (CLAS)
will be upgraded to the CLAS12 detector (Fig.\,\ref{CD3_CLAS12}) to meet the requirements of the study of the structure
of nucleons and nuclei. Its main features include operating at a luminosity of 10$^{35} cm^{-2} s^{-1}$ , 
a ten-fold increase over the current CLAS operating conditions, and
detection capabilities and particle identification for forward-going high momentum charged
and neutral particles.
The Hall C facility will use the 
existing High Momentum Spectrometer (HMS) together with 
a new Super High Momentum Spectrometer (SHMS), powerful enough to analyze
charged particles approaching the beam momentum. The SHMS will cover a solid angle up to 
4 msr with sensitivity down to  5.5$^{\circ}$ using a 
small horizontal-bend magnet.
The beamline into Hall A will be upgraded to
transport the full 11 GeV beam into the experimental area for experiments using the existing
spectrometers, and for installation of major specialized experiments. 
The  Hall D facility (Fig.\,\ref{Photon_beam_layout})
will use the 12 GeV electron beam to produce a coherent 
bremsstrahlung beam. This photon beam, which peaks  in the energy range between 8.4 and 9.0 GeV,
is 40\% linearly polarized and will be used to carry out a program in 
gluonic spectroscopy with the  new Hall D detector (Fig.\,\ref{hddetector_platform}).

\section{Gluonic excitations}
The observation, nearly four decades ago, that mesons are grouped in nonets, each characterized
by  unique values of $J^{PC}$ --
spin ($J$), parity ($P$) and charge conjugation ($C$) quantum
numbers -- led to the development of the  quark model.  Within this picture, mesons
are bound states of a quark ($q$) and antiquark ($\bar q$).  
The three light-quark flavors
($up$, $down$ and $strange$) suffice to explain the spectroscopy of most  
-- but not all -- of the lighter-mass mesons (below 3~GeV/c$^2$).
Early observations yielded only those $J^{PC}$ quantum numbers consistent
with a fermion-antifermion bound state.
Other $J^{PC}$ combinations, such as $0^{--}$, $0^{+-}$,
$1^{-+}$ and $2^{+-}$, require additional degrees of freedom and are called \emph{exotic} in this context.

Our understanding of how quarks form mesons has evolved within quantum chromodynamics 
(QCD) and we expect a rich spectrum of mesons that takes into account not
only the quark degrees of freedom, but also the gluonic degrees of freedom. 
Excitations of the gluonic field binding the quarks can give rise to 
so-called \emph{hybrid} mesons. (For a review see Ref.\,\cite{Klempt:2007cp}). A picture of these hybrid mesons
is one where these particles are excitations of a gluonic flux tube 
that forms between the quark and antiquark.   Particularly interesting 
is that many of these hybrid mesons are expected to have exotic $J^{PC}$ quantum numbers,
which simplifies the spectroscopy because they do not mix with
conventional  $q\bar{q}$ states.  
The level splitting between the ground state flux tube and the first excited
transverse modes is expected to be
about 1~GeV/c$^2$, and lattice QCD calculations \cite{Bernard:2003jd}  indicate
the lightest exotic hybrid (the $J^{PC} = 1^{-+}$) has a mass of about $1.9$~GeV/c$^2$.
The \GX\ experimental search in Hall D has a mass reach up to about $2.8$~GeV/c$^{2}$ 
to observe mesons with masses up to 2.5~GeV/c$^{2}$.

\begin{figure}[b]
\begin{minipage}[t]{=7.5cm}
\begin{center}
\includegraphics[width=7cm,bb=15 15 620 435]{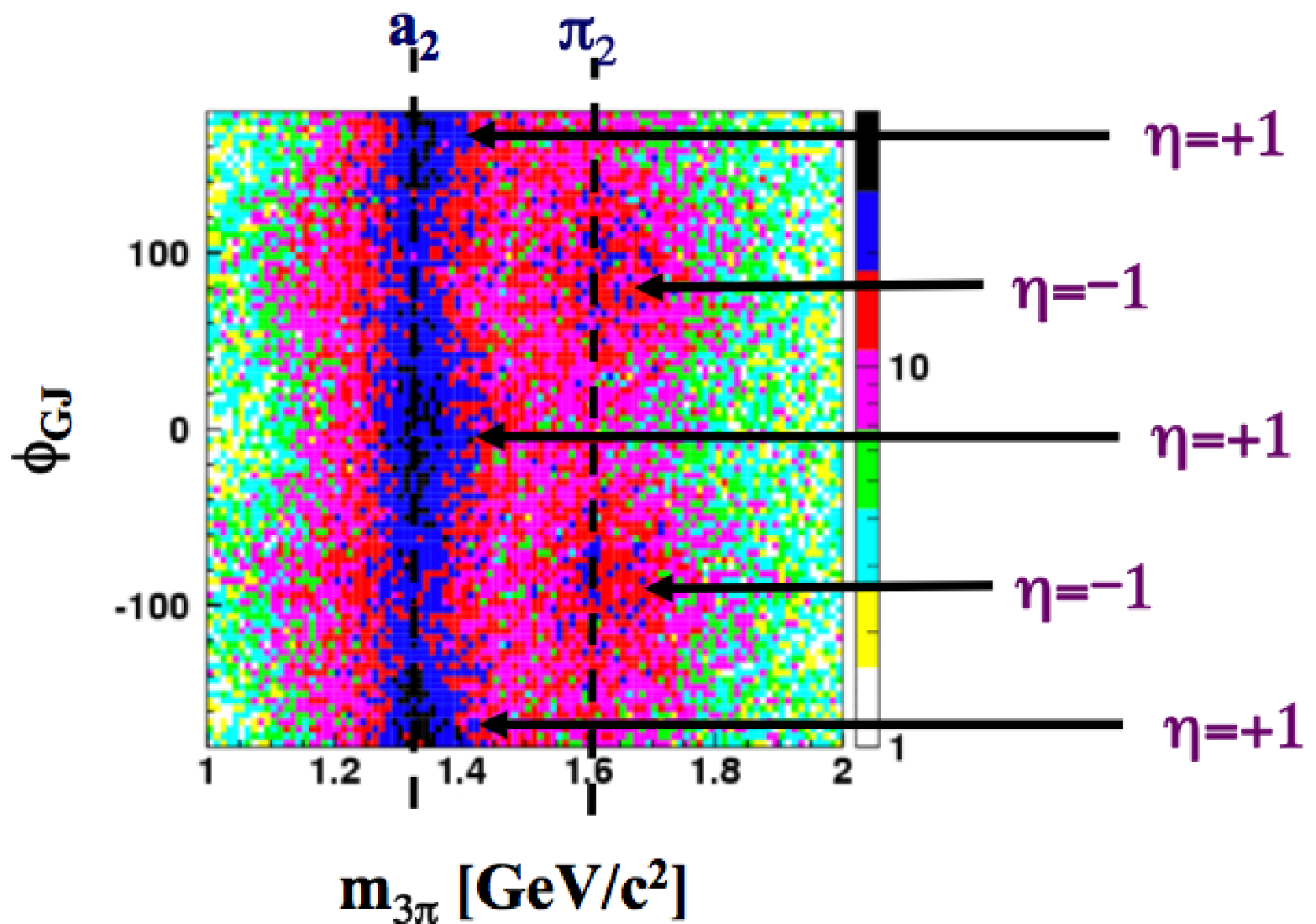}
\caption{\label{ang_distr} Correlations between the angular decay distributions
and the mass of the parent system decaying to three pions. The major particles in
the spectrum are indicated as well as nodes in the decay spectrum, highlighting the
wealth of information in decay correlations when the incident photon is polarized.}
\end{center}
\end{minipage}
\hfill
\begin{minipage}[t]{5cm}
\begin{center}
\includegraphics[width=4.5cm]{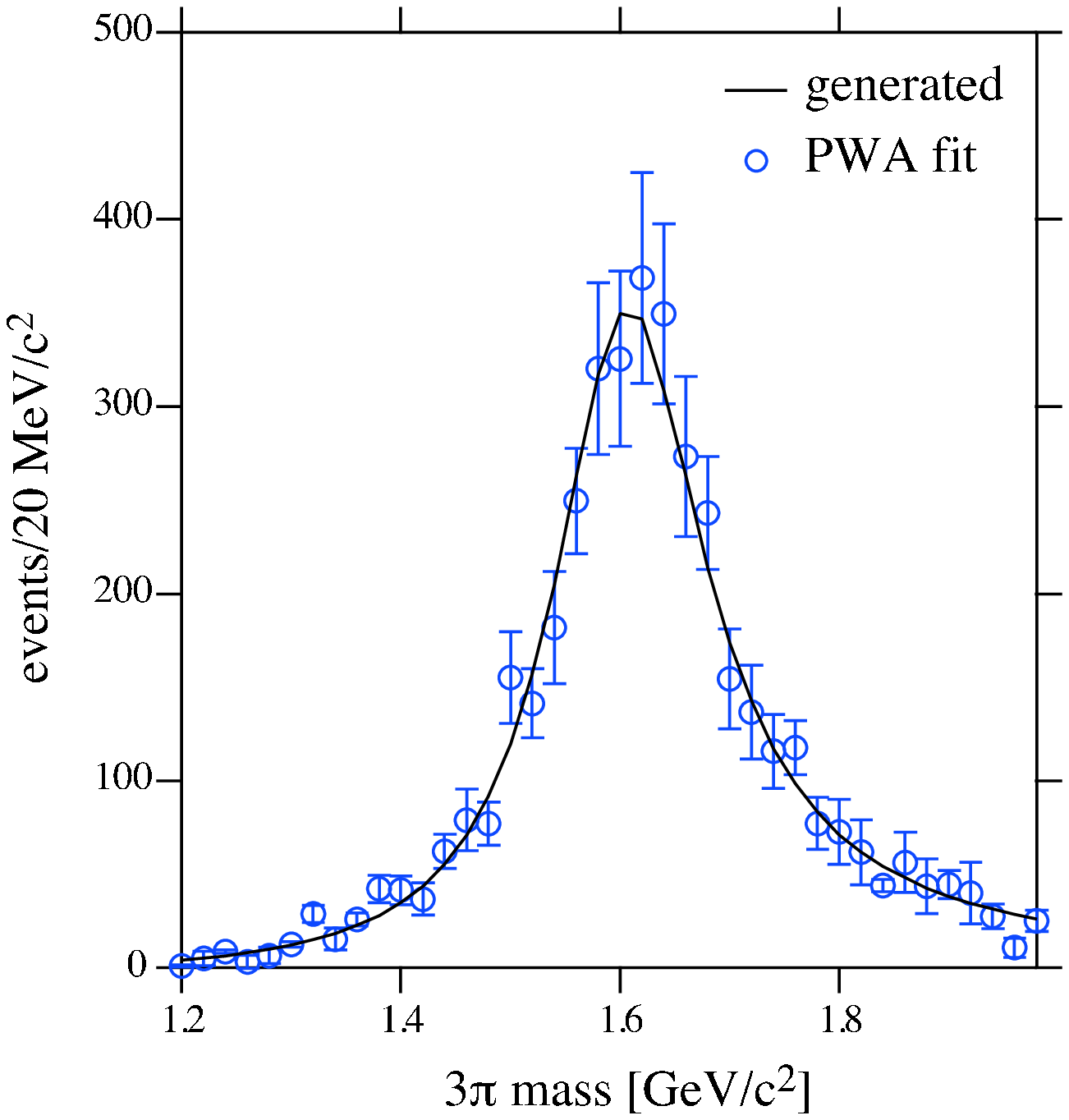}
\caption{\label{ch11_blindpwa} Expected precision for extracting an exotic
wave representing 2.5\% level of the total sample consisting of six non-exotic waves.}
\end{center}
\end{minipage}
\end{figure}

There are tantalizing suggestions, mainly from experiments using beams of $\pi$ mesons,
that exotic hybrid mesons do exist.  The evidence is by no means  clear cut, owing in part, 
to the apparently small production rates for these states in the decay channels examined.
It is safe to conclude that the extensive data collected to date with $\pi$ probes have 
not  uncovered the hybrid meson spectrum. 
Based on models, such as the flux-tube model, we expect the production of hybrid mesons
in photon induced reactions to be comparable to the production of normal mesons. 

Photoproduction of mesons using an $\approx 9$~GeV, linearly polarized photon beam provides
a unique opportunity to search for exotic hybrids. Existing data is extremely limited for
charged final states, and no data exist for multi-neutral final states. To carry out
such a search, \GX\ will need to look at many different final states involving both
charged particles and photons, but particular emphasis will be placed on those reactions
that have $3$ or more pions in the final state. The discovery potential for \GX\ comes
first from the very high statistics based on $10^{7}$ tagged $\gamma$/s
on target, which will exceed existing photoproduction data by  $4$ to $5$ orders of 
magnitude.
Second, \GX\ has the ability to study many different 
final states in the same detector. These two capabilities  will identify hybrids, if
they exist at the few percent level, and also map out their decay properties. 

Determining the quantum numbers of mesons produced in the \GX\ experiment will 
require an amplitude analysis based on measuring the energy and momentum
of their decay products. An example of the rich information content in the decay angular distributions
is shown in Fig.\,\ref{ang_distr}, which is exploited in the amplitude analysis. 
In a partial-wave analysis exercise on Monte Carlo data, exotic waves of order
a few percent of the total sample could be extracted reliably as shown in Fig.\,\ref{ch11_blindpwa}.
In summary, the \GX\ detector has been designed to carry out a broad program
to study gluonic excitations and provide extensive data to search for
exotic mesons  in the essentially unexplored territory of photoproduction reactions.

\section{3D view of the nucleon}

\begin{figure}[b]
\begin{minipage}[t]{=7.5cm}
\begin{center}
\includegraphics[width=7cm,bb=16 35 600 300]{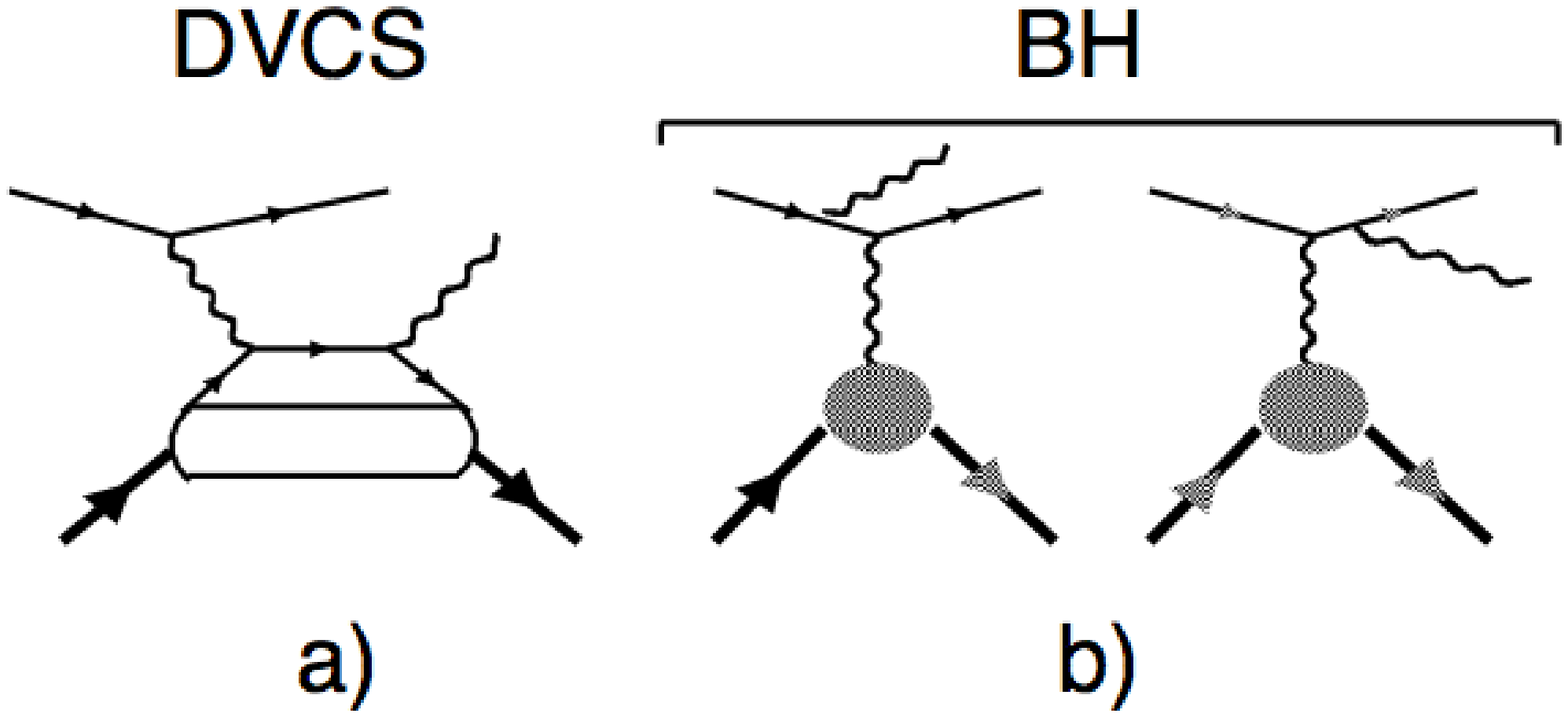}
\caption{\label{dvcs_fdia} Schematic handbag diagrams for deeply
virtual Compton scattering (left) and Betler-Heitler process (right).}
\end{center}
\end{minipage}
\hfill
\begin{minipage}[t]{5.5cm}
\begin{center}
\includegraphics[width=5.5cm]{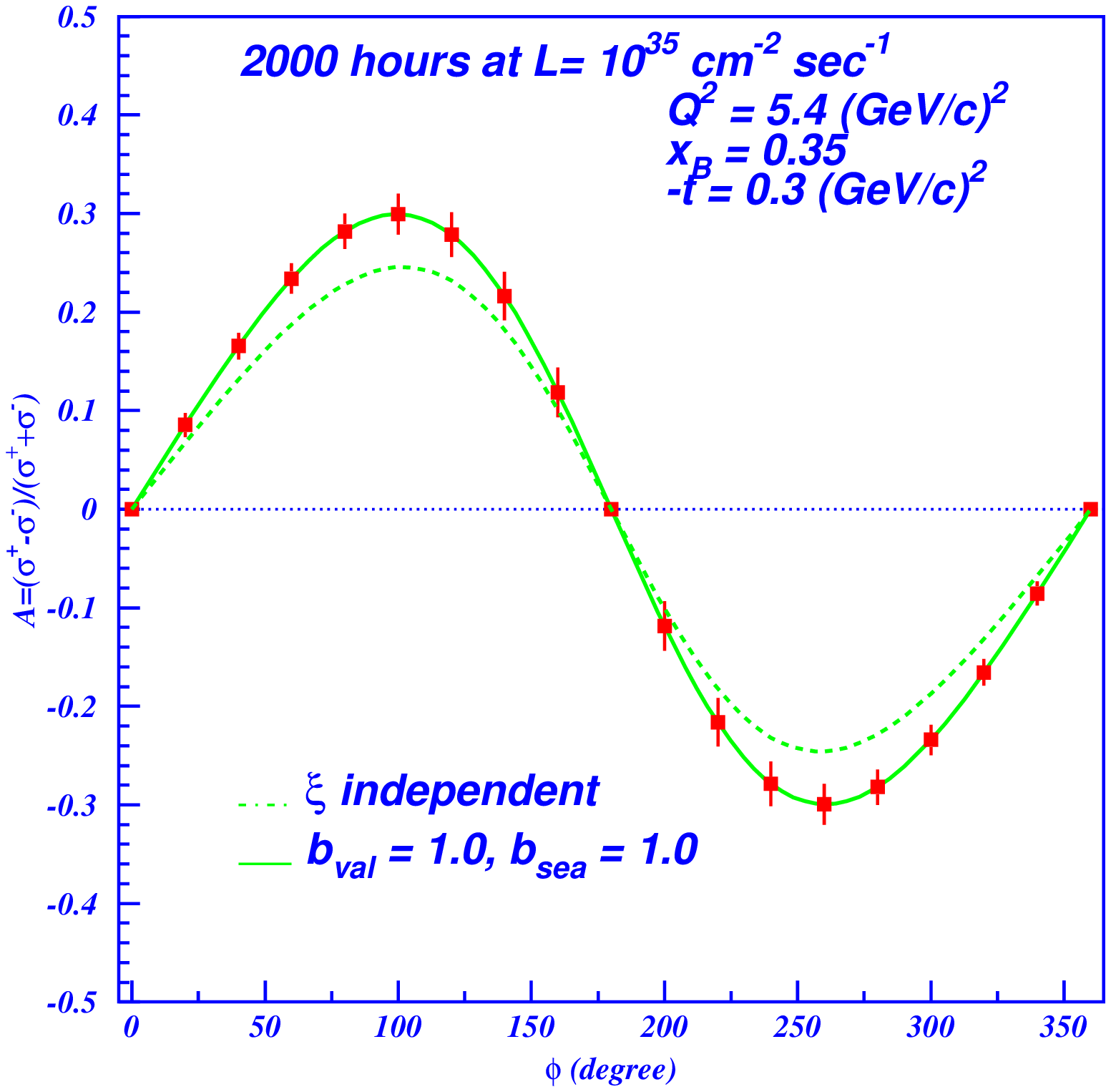}
\caption{\label{drawplotu} Expected precision in the measurement of a
DVCS asymmetry as a function of the azimuthal angle $\phi$
for a specific kinematic point specified by $Q^2, x_B$ and $-t$. 
The curves show typical predicted model dependencies of the asymmetry.}
\end{center}
\end{minipage}
\end{figure}

Historically, electron scattering experiments have focussed either 
on the measurements of form factors, using exclusive processes, or on measurements
of inclusive processes to extract deep inelastic structure functions.
Elastic processes measure the momentum transfer dependence of the form factors,
while the latter ones probe the quark's longitudinal momentum
and helicity distributions in the infinite momentum frame.
Form factors and deep inelastic structure functions measure two different
slices of the proton structure. While it is clear that the two 
pictures must be connected, a common framework for the interpretation
of these data has only recently been developed using  
Generalized Parton Distribution (GPD) functions.
Mapping out the GPD's will allow, for the first time, to 
obtain a 3-dimensional picture of the nucleon. For reviews of this subject, see e.g. Refs.\,\cite{Diehl:2003ny,Belitsky:2005qn}.

Unravelling the information about GPDs from the data is not a simple task.
It requires an extensive experimental program and detailed analysis with
controlled theoretical corrections. 
Measurements of cross sections on the proton and neutron, as well as beam
spin asymmetry measurements, will be needed
to disentangle the GPDs. The reaction $ep \rightarrow ep \gamma$ (Fig.\,\ref{dvcs_fdia})
includes the physically interesting amplitude for 
Deeply virtual Compton scattering (DVCS) . This process 
is the cleanest tool for constraining GPDs from the
data, because it is in the most advanced stage of theoretical studies.
Therefore we use DVCS to illustrate the wealth of data which will be accessible 
using CLAS12. 
The reaction is dominated by the
Bethe-Heitler (BH) amplitudes, but the interference term between DVCS and BH can be probed using
polarized electron beams via the single beam spin asymmetry, which is
dominated by the $\sin{\phi_{\gamma \gamma^*} }$ moment.
In this case, the small DVCS amplitude which depends on the GPD's, is 
amplified by the larger but well-known BH amplitude.


The large coverage for photons by the CLAS12 detector will allow clean identification of
the DVCS events, and with the large acceptance,
cross sections and spin asymmetries will be
measured in a large number of kinematic bins simultaneously.
From the more than one thousand kinematic bins measured, we show one example of
the beam spin asymmetry in Fig.\,\ref{drawplotu}. 
The expected experimental precision
is sufficient to be sensitive to different model calculations which are also shown on the figure.

\section{Status of the project and summary}
The  approval for construction of the project, known as Critical Decision 3,  was received in September 2008.
Construction funding has begun, with commissioning planned to start in 2013. 
Contracts for civil construction are in progress and orders for major components for the
accelerator and experimental equipment are being placed.
We have presented two examples of the exciting physics program envisioned for this facility,
demonstrating some of its unique features that will extend our understanding of the
strong interactions. 

\section{Acknowledgments}
This work was supported by the U.S. Department of Energy contract DE-AC05-06OR23177, under which 
Jefferson Science Associates, LLC operates the Thomas Jefferson National Accelerator Facility. I would 
like to thank my colleagues from the \GX\ collaboration and 
the JLab Upgrade project, especially E. Aschenauer and L. Elouadrhiri,
for useful comments and help in preparing materials for this report.

%
%
%
%
%
%
%
%
\bibliographystyle{unsrt}
                                                                                        
\bibliography{panic08}

\hyphenation{Post-Script Sprin-ger}
\begin{thebibliography}{1}

\bibitem{12gevCDR}
{Conceptual Design for the Science and Experimental Equipment for the 12 GeV
  Upgrade of CEBAF}.
\newblock
  http://www.jlab.org/div\_dept/physics\_division/GeV/doe\_review/CDR\_for\_Sc%
ience\_Review.pdf, 2005.

\bibitem{Klempt:2007cp}
Eberhard Klempt and Alexander Zaitsev.
\newblock {Glueballs, Hybrids, Multiquarks. Experimental facts versus QCD
  inspired concepts}.
\newblock {\em Phys. Rept.}, 454:1--202, 2007.

\bibitem{Bernard:2003jd}
C.~Bernard et~al.
\newblock {Lattice calculation of 1-+ hybrid mesons with improved
  Kogut-Susskind fermions}.
\newblock {\em Phys. Rev.}, D68:074505, 2003.

\bibitem{Diehl:2003ny}
M.~Diehl.
\newblock {Generalized parton distributions}.
\newblock {\em Phys. Rept.}, 388:41--277, 2003.

\bibitem{Belitsky:2005qn}
A.~V. Belitsky and A.~V. Radyushkin.
\newblock {Unraveling hadron structure with generalized parton distributions}.
\newblock {\em Phys. Rept.}, 418:1--387, 2005.

\end{thebibliography}

\end{document}